\documentclass[twocolumn]{aastex61}
\pdfoutput=1 
\usepackage{amsmath,amstext}
\usepackage[T1]{fontenc}
\usepackage{apjfonts}
\usepackage[figure,figure*]{hypcap}
\usepackage{longtable}
\usepackage{comment}
\usepackage{bmpsize}

\received{}
\revised{}
\accepted{}
\submitjournal{ApJL}

\shorttitle{B/T - j/M relation}
\shortauthors{Sweet et al.}

\begin{document}

\title{Revisiting the stellar mass -- angular momentum -- morphology relation: extension to higher bulge fraction, and the effect of bulge type.}

\correspondingauthor{Sarah Sweet}
\email{sarah@sarahsweet.com.au}

\author{Sarah M. Sweet}
\author{David Fisher}
\author{Karl Glazebrook}
\affiliation{Swinburne}
\author{Danail Obreschkow}
\author{Claudia Lagos}
\author{Liang Wang}
\affiliation{UWA}

\begin{abstract}
We present the relation between stellar specific angular momentum $j_*$, stellar mass $M_*$, and bulge-to-total light ratio $\beta$ for THINGS, CALIFA and Romanowsky \& Fall datasets, exploring the existence of a fundamental plane between these parameters as first suggested by Obreschkow \& Glazebrook. Our best-fit $M_*-j_*$ relation yields a slope of $\alpha = 1.03 \pm 0.11$ {with a trivariate fit including $\beta$}. When ignoring the effect of $\beta$, the exponent $\alpha = 0.56 \pm 0.06$ is consistent with $\alpha = 2/3$ predicted for dark matter halos.
There is a linear $\beta - j_*/M_*$ relation for $\beta \lesssim 0.4$, exhibiting a general trend of increasing $\beta$ with decreasing $j_*/M_*$. Galaxies with $\beta \gtrsim 0.4$ have higher $j_*$ than predicted by the relation. Pseudobulge galaxies have preferentially lower $\beta$ for a given $j_*/M_*$ than galaxies that contain classical bulges. Pseudobulge galaxies follow a well-defined track in $\beta - j_*/M_*$ space, consistent with Obreschkow \& Glazebrook, while galaxies with classical bulges do not. These results are consistent with the hypothesis that while growth in either bulge type is linked to a decrease in $j_*/M_*$, the mechanisms 
that build pseudobulges seem to be less efficient at increasing bulge mass per decrease in specific angular momentum than those 
that build classical bulges.
\end{abstract}


\section{Introduction}
Galaxy stellar mass $M_*$ and angular momentum $J$ are fundamental properties of galaxies: they have been shown to correlate strongly with 
 galaxy size and density \citep{Mo+1998}, disk thickness and colour \citep{Hernandez+2006}, and morphology \citep[][hereafter RF12, OG14, {C16}]{RF12,OG14,Cortese+2016}. 

$M$ and $J$ are not independent in that $J$ is scaled by mass, so the standard method for studying their relationship is to remove the mass dependence to obtain specific angular momentum $j = J/M$. Specific angular momentum of baryons in the galaxy $j_{baryons}$ is empirically similar to that of the dark matter halo $j_h$ \citep{Fall1983}. This similarity is expected for baryons in the halo, since the same tidal forces are experienced during spin-up \citep{Catelan+1996a,Catelan+1996b,vandenBosch+2001,Barnes+1987}, but has been historically difficult to reconcile for baryons in the disk, \emph{viz.} the ``angular momentum catastrophe'' \citep{Governato+2010,Agertz+2011}. $j_{baryons}$ is typically further resolved into analogous specific angular momenta for stars, H$\alpha$, HI and H$_2$ ($j_*$, $j_{H\alpha}$, $j_{HI}$, $j_{H_2}$ respectively) depending on the observed kinematics and mass profiles\footnote{Surface density is typically used as a proxy; {$j$ does not have the mass scaling of $J$}, but the mass profile is used as a weighting factor.} available to be studied. In this work we focus on stellar specific angular momentum $j_* = J_*/M_*$.

The observational $M_*-j_*$ plane was first studied by \citet{Fall1983}, who found $j_* \propto qM_*^\alpha$, with {parallel tracks defined by late-type and early-type galaxies. These tracks have exponent $\alpha = 2/3$, in agreement with the natural scaling of CDM halos in a hierarchical universe. 
 \citet[][RF12]{RF12} later studied the relation between $M_*$, $j_*$ and $\beta$, confirming the earlier result of $j \propto M_*^{2/3}$, with factor $q$ depending on whether disks or bulges are considered.}
\citet[][OG14]{OG14} used high-quality observations of 16 galaxies from The HI Nearby Galaxy Survey \citep[THINGS,][]{Leroy+2008,Walter+2008} to further investigate the $M_*-j_*-\beta$ relation, finding that $\alpha = 2/3$ for $0 \leqslant \beta \leqslant 0.32$, but that $\alpha \sim 1$ {when $\beta$ is treated as a free parameter.}
{More recently, C16 analysed a subset of galaxies from the Sydney-AAO Multi-object Integral field \citep[SAMI,][]{Croom+2012} Galaxy Survey \citep{Bryant+2015,Allen+2015,Sharp+2015}, and similarly found, when considering $j$ at one effective radius $r_e$, that $\alpha$ is consistent with 2/3 for the whole range of morphologies, but higher when a single morphology class is considered, and approaching $\alpha = 1$ for late-type galaxies.}

There is a known dichotomy in the properties of pseudo- vs. classical bulges \citep[e.g.][]{kormendy2004,fisher2016}; classical bulges are pressure-supported components thought to form {by minor mergers \citep{Toomre1977,Schweizer1990} or disk instabilities \citep{Toomre1964}}, while pseudobulges are rotationally-supported components formed during secular evolution of the disk, so naturally have higher $j_*$ \citep{kormendy2004,Wyse+1997}. Classical bulges generally contribute a larger $\beta$ than pseudobulges \citep{fisher2016}. These properties are intimately related to galaxy angular momentum and morphology. However, previous studies of the $M_*-j_*-\beta$ relation have not analysed {galaxies that contain} classical bulges separately from {those that contain} pseudobulges, so the effect of bulge type is unknown.

In this work we investigate the effect of bulge type on the relationship between stellar mass, specific angular momentum and morphology across a large range of $\beta$, by combining OG14 with high-quality subsets of the sample in RF12 and the Calar Alto Legacy Integral Field Area Survey \citep[CALIFA,][]{Sanchez+2012,Husemann+2013,Walcher+2014,Sanchez+2016}. 
In Section~\ref{sec:methods} we describe our methods for measuring bulge properties and $j_*$, and introduce the datasets. In Section~\ref{sec:results} we present the $M_* - j_* - \beta$ relation as it relates to bulge type, given the known dichotomy in the properties of pseudo- vs. classical bulges {and the galaxies that host them} \citep[e.g.][]{kormendy2004,fisher2016}. Section~\ref{sec:conclusion} concludes this letter with a discussion of the significance of these results. 

\section{Sample \& Methods\label{sec:methods}}
We combine observations from THINGS, RF12 and CALIFA datasets to trace the fundamental relation between $M_*$, $j_*$ and $\beta$ over a wide range of $\beta$. The three samples are complementary. We have high-quality $j_*$, $\beta$ and bulge classifications for THINGS, but the sample is limited to low to moderate $\beta$, with few {galaxies that contain} classical bulges. We thus employ the RF12 galaxies for which we have high-quality $\beta$ and bulge classification to extend our sample to higher $\beta$ and increase the number of {galaxies with} classical bulges. Similarly, we also include a subset of the CALIFA sample for which we measure high-quality $j_*$, and a greater range of $\beta$. Below we present our methods for determining bulge properties and $j_*$, before giving specific details for each of our samples.

\subsection{Bulge-to-total mass ratio and type}
\begin{figure}
\plotone{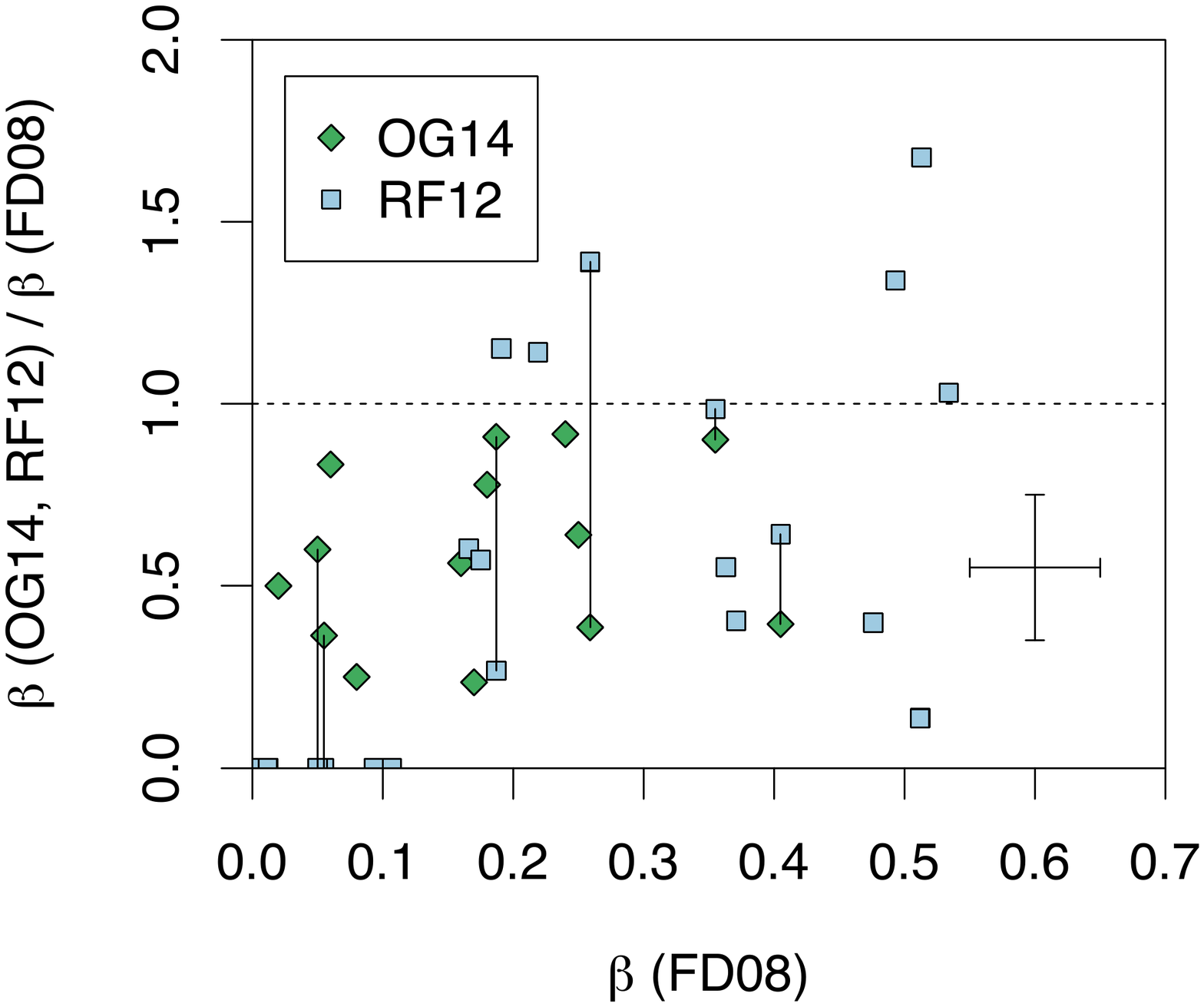}
\caption{Comparison between bulge-to-total ratio $\beta$ methods. {THINGS galaxies presented in OG14 are shown as green filled diamonds; galaxies in RF12 as blue filled squares. }
Galaxies appearing in both RF12 and OG14 are joined with solid lines. The dashed line represents the 1:1 relation. A typical error bar is shown. Both RF12 and OG14 methods {tend to} underestimate $\beta$ with respect to our method described in \citet{fisher2008}. 
\label{fig:betabeta}}
\end{figure}

We obtain bulge properties by cross-correlating the OG14 and RF12 samples with the combined data set of \citet{fisher2010,fisher2011}, \citet{fabricius2012} and \citet{fisher2013}.  These samples use the same method, software and wavelength range to conduct 2D bulge-disk decompositions \citep[described in][FD08]{fisher2008}. The method combines high-resolution HST imaging with wide-field ground based imaging to reduce uncertainties and degeneracies inherent to bulge-disk decompositions. It also accounts for the different mass-to-light ratio of the bulge and disk.  Conversely, OG14 simply measured the bulge as the excess light over fitted exponential disk, while RF12 fitted two elliptical isophotes in projection \citep{Kent1986}. Figure~\ref{fig:betabeta} compares FD08 $\beta$ with OG14 and RF12, illustrating that OG14 and RF12 present $\beta$ that are mutually inconsistent and systematically underestimated with respect to FD08. Importantly, our consistent method allows for an accurate comparison of bulge properties between OG14 and RF12 {, with an uncertainty of $\Delta \beta_{FD08} = \pm 0.05$.}

We use the well-known correlation between bulge S\'ersic index $n_b$ \citep{Sersic1963} and bulge type \citep{fisher2008,fisher2016} to classify {galaxies that contain} pseudobulges and classical bulges, such that $n_b<2$ implies a {galaxy with a} pseudobulge and $n_b>2$ implies a {galaxy with a} classical bulge. The exception is NGC3593 in RF12, with a low $n_b\approx1.2$, which would ordinarily imply {that it contains} a pseudobulge. However, its bulge is ``not classifiable'' \citep{fabricius2012}, for the following reasons: 1) NGC3593 is an extreme example of counter-rotating kinematics \citep{fabricius2012,bertola1996}, suggesting a recent merger \cite[e.g.][]{bassett2017}, and the empirical methods of bulge classification fail for most galaxies that are experiencing interactions \citep{fisher2016}; 2) the surface photometry cannot be reliably fit due to chaotic dust profile \citep{ravindranath2001}; 3) the galaxy has a ``peculiar'' global morphology \citep[e.g.][]{sandage1994}. 

\subsection{Specific angular momentum}
We adopt the same method as \citet{Obreschkow+2015} for calculating $j_*$ from integral field spectroscopic observations (IFS). Compared with slit spectroscopy, this method significantly increases accuracy in tracing the kinematic field, since kinematic and photometric major axes may be misaligned \citep[e.g.][]{Sweet+2016}, and the velocity fields of many galaxies are not well described by simple 1D rotation curves. 

{We use a combination of the observed deprojected angular momentum where available, together with {a model-informed estimate of the} deprojected angular momentum in the spaxels where observations are not of sufficient quality, e.g. due to low signal-to-noise ratios {\bf $< 3$} in either the stellar surface density maps or the kinematic maps. }

1)  The observed {deprojected} angular momentum $J_i = {\bf r}_i {\bf v}_i m_i$ in every spaxel $i$ at deprojected radius ${\bf r}$ whose circular velocity ${\bf v}$ is derived from kinematic maps and mass $m$ from stellar surface density maps. {The deprojection is based on inclination and position angle derived from a fit to the stellar surface density maps; inclination and position angle are assumed to be constant with radius. Non-circular motions are neglected in this work, but see Sweet et al. (in prep.) for a treatment of the contribution of non-circular motions to total and spatially-resolved $j_*$.}

2) The model ${J}_i$ at each spaxel is computed by fitting an exponential profile to the disk, in order to reach the total angular momentum, traced by the flat part of the rotation curve. {The surface mass density is characterised by $\Sigma({\bf r}_i) \propto \rm{exp}(-{\bf x}_i)$, where ${\bf x}_i = {\bf r}_i/r_{flat}$ and {the exponential scale length} $r_{flat}$ is {\bf assumed to be} the radius at which the velocity reaches the converged velocity $v_{flat}$. {While not the case in general, this simplifying assumption is made in order to keep the number of free parameters at a minimum.}} {The exponential fit is given by equation 7 of OG14; ${\bf \tilde{v}}_i \approx v_{flat} \left(1 - \rm{exp}(-{\bf x}_i)\right)$. Following OG14 equation 8, the model ${J}_i = 2((1+{\bf x}_i)^3-1)/(1+{\bf x}_i)^3 {\bf r}_i v_{flat} m_i$.} {The model is on average consistent with the observed $J$ to the 5\% level, if both are summed over the same high signal-to-noise spaxels. However, note that the purpose of the model is only to serve as an estimate of $J_i$ in the low signal-to-noise spaxels. }

3) The total $j_*$ is then given by $J/M_*$, where $J = \lvert\sum^i {J}_i\rvert$ is the norm of the sum over the observed $J_i$ where defined, and {estimated} $J_i$ in other spaxels,  {integrated to ${\bf r}_i=\infty$}. {Including the estimated $J_i$ in the spaxels where data is missing comprises an average of 20\% of the total $j_*$. }

{The uncertainty in this method is typically $\Delta j_* / j_* = \pm$ 5-10\%, predominantly contributed by {the uncertainty in extrapolating $j_*(r)$ beyond where it is converged} (that is, {the assumption} that the observations reach the flat part of the rotation curve), as well as the assumption of pure circular motion, and the uncertainty in inclination (see OG14 for further details).}

\subsection{THINGS}
THINGS \citep{Walter+2008} is a survey of 34 nearby galaxies observed to high multiples of effective radius $r_e$. OG14 presented $j_*$ measured with HI kinematics for the sixteen spiral THINGS galaxies that have stellar surface density maps published by \citet{Leroy+2008}. {Stellar masses and uncertainty $\Delta M_*/M_* = \pm 0.11$ dex are also taken from OG14.}
This sample contains 13 {galaxies with} pseudobulges and 3 {with} classical bulges, with bulge-to-total mass ratios $0 \leqslant \beta \leqslant 0.41$.

\subsection{Romanowsky \& Fall}
RF12 presented $j_*$ for a sample of spiral and elliptical galaxies, calculated using stellar kinematics from slit spectroscopy of starlight and ionized gas. OG14 found these to systematically {vary} with respect to their own IFS observations, so we rescale RF12 $j_*$ {using equation 6 of OG14:
\begin{equation}
\label{eqn:OG14-6}
\left(\frac{j_*}{10^3\ {\rm kpc\ km\ s^{-1}}}\right) \approx 
 1.01\left(\frac{\widetilde{j_*}}{10^3\ {\rm kpc\ km\ s^{-1}}}\right)^{1.3}
\end{equation}}
{The relative uncertainty is $\Delta j_* / j_* = \pm 32\%$, given by the quadrature sum of the uncertainty in RF12 $j_*$ (10\%) and the RMS scatter of the calibrating relation (30\%).}

{RF12 derived $M_*$ using \citet{Bell+2003} colours and a diet Salpeter IMF located between \citet{Kroupa2001} and \citet{Salpeter1955}, 
which translates to an assumed $K-$band mass-to light ratio $M/L{_K} = 1M_\odot/L_{\odot,K}$. This differs to \citet{Leroy+2008}, who assumed a \citet{Kroupa2001} IMF and consequently $M/L{_K} = 0.5M_\odot/L_{\odot,K}$. We therefore scale the RF12 $M_*$ by 0.5 to achieve consistency with our THINGS sample. The uncertainty in RF12 $M_*$ of $\Delta M_*/M_* = \pm 0.2$ dex is taken from OG14.}

Motivated by the desire to calculate $\beta$ and classify bulge type in the same manner as for our THINGS sample, we select the 25 galaxies from RF12 for which we have existing high-quality bulge-disk decompositions. 
The RF12 sample thus contains 12 {galaxies that contain} pseudobulges, 12 {that contain} classical bulges and one {whose bulge is} unclassifiable (NGC3593), with $0 \leqslant \beta \leqslant 0.53$.

\subsection{CALIFA}
The CALIFA survey has made available stellar kinematic and surface density maps for 300 nearby galaxies \citep{Falcon-Barroso+2017}. We use the OG14 method described above to measure $j_*$ for these galaxies, and note that $j_*(<r)$ converges to $>0.99j_*$ at a radius $r \sim 3r_e$. This motivates us to select the subset observed to at least that radius. 

We take stellar masses from \citet{Falcon-Barroso+2017}{, who used the methods outlined in \citet{Walcher+2014}; namely, assuming \citet{Bruzual+2003} stellar populations and a \citet{Chabrier2003} IMF. The RMS scatter between their two implementations gives the uncertainty $\Delta M_*/M_* = \pm 0.15$ dex. The \citet{Bruzual+2003} stellar populations were shown by \citet{Sanchez+2013} to give stellar masses consistent with those derived from \citet{Bell+2001} colours (which themselves are consistent with \citealt{Bell+2003}, as used by \citealt{Leroy+2008}). 
$K$-band $M/L{_K}$ ratios based on the \citet{Chabrier2003} IMF differ from those assuming a \citet{Kroupa2001} IMF \citep[as in][]{Leroy+2008} by only 10\% \citep[][table 2]{Longhetti+2009}, well within the scatter. Hence, we are comfortable that the CALIFA stellar masses are comparable with those of our THINGS sample.}

 Hubble types are taken from \citet{Falcon-Barroso+2017} and $\beta$ from the bulge-disk decompositions presented in \citet{Mendez-Abreu+2017}. \citet{Mendez-Abreu+2017} includes only one FD08 galaxy; $\beta_{FD08} = 0.5$ \emph{cf.} $\beta_{CALIFA} = 0.6$. {Since the \citet{Mendez-Abreu+2017} decompositions were based on SDSS imaging, we estimate the uncertainty as $\Delta \beta_{CALIFA} = \pm 0.1$}. We remove five pure elliptical galaxies, since $j_*$ as measured here is strictly applicable to systems that contain a disk. There is a lack of imaging of sufficient resolution to reliably recover the bulge S\'ersic index, so we do not categorise CALIFA {galaxies} into {those with} classical {or} pseudobulges.  Our high-quality CALIFA subset comprises 35 spiral and 15 elliptical/lenticular galaxies, and spans $0 \leqslant \beta \leqslant 0.73$.\\

The properties of the resulting samples are given in Table~\ref{tab:data}.

\begin{table*}
\centering
\caption{Properties of galaxies presented in this paper.} 
\label{tab:data}
\begin{tabular}{lllrrrrrrrrrrrr}
  \hline
Name & Survey & Type & $M_*$ & $\Delta M_*/M_*$ & $\beta$ & $\Delta \beta$ & $r_d$ & $r_{flat}$ & $v_{flat}$ & $j_*$ & $\Delta j_*$ & $n_{bulge}$ \\ 
  &   &   & [log($M_\odot$)] & [dex]&   &   & [kpc] & [kpc] & [km s$^{-1}$] & [kpc km s$^{-1}$] & [kpc km s$^{-1}$] &\\ 
(1) & (2) & (3) & (4) & (5) & (6) & (7) & (8) & (9) & (10) & (11) & (12) & (13) \\
  \hline
NGC0628 & THINGS & Sc & 10.10 & 0.11 & 0.17 & 0.05 & 2.3 & 0.8 & 217 & 955 & 95 & 1.53 \\ 
  NGC0925 & THINGS & SBcd & 9.90 & 0.11 & 0.06 & 0.05 & 4.1 & 6.5 & 136 & 871 & 87 & 0.90 \\ 
  NGC2403 & THINGS & SBc & 9.70 & 0.11 & 0.06 & 0.05 & 1.6 & 1.7 & 134 & 417 & 42 & 0.80 \\ 
  ... & ... & ...& ... & ... & ... & ... & ... & ... & ... & ... & ... & ... \\
  NGC0224 & RF12 & Sb & 10.76 & 0.20 & 0.48 & 0.05 & 5.9 &  & 234 & 2967 & 938 & 2.13 \\ 
  NGC0247 & RF12 & Sd & 9.54 & 0.20 & 0.00 & 0.05 & 4.1 &  & 92 & 749 & 237 & 0.00 \\ 
  NGC0300 & RF12 & Sd & 8.93 & 0.20 & 0.00 & 0.05 & 1.6 &  & 60 & 173 & 55 & 1.64 \\ 
  ... & ... & ... & ... & ... & ... & ... & ... & ... & ... & ... & ... & ... \\
  IC1151 & CALIFA & Scd & 9.85 & 0.15 & 0.02 & 0.10 & 1.9 & 4.0 & 113 & 1122 & 154 &  \\ 
  MCG-02-02-030 & CALIFA & Sb & 10.37 & 0.15 & 0.08 & 0.10 & 3.1 & 8.8 & 177 & 2698 & 340 &  \\ 
  NGC0001 & CALIFA & Sbc & 10.80 & 0.15 & 0.46 & 0.10 & 1.8 & 3.4 & 169 & 1564 & 117 &  \\ 
  ... & ... & ... & ... & ... & ... & ... & ... & ... & ... & ... & ... & ... \\
   \hline
\end{tabular}\\
\smallskip
{Columns: (1) galaxy identifier; (2) dataset; (3) Hubble type; (4) log(stellar mass); (5) measurement uncertainty in $M_*$; (6) bulge-to-total ratio; (7) measurement uncertainty in $\beta$; (8) scale length; (9) radius at which rotation curve becomes flat; (10) asymptotic velocity; (11) stellar specific angular momentum; (12) measurement uncertainty in $j_*$; (13) bulge S{\'e}rsic index. 
A machine-readable version of this table is available online. A portion is shown here for guidance regarding its form and content.}
\end{table*}

\section{The relation between stellar mass, specific angular momentum and morphology\label{sec:results}}

We present our $M_*-j_*-\beta$ relation for THINGS, RF12 and CALIFA, fitting the data with the log-linear three-parameter model given in eq.~(9) of OG14,
\begin{equation}\label{eq:master}
	\beta = p_1\log_{10}M_*+p_2\log_{10}j_*+p_3,
\end{equation}
where $\beta$ is the bulge-to-total stellar mass ratio, $M_*$ is the stellar mass in units of $10^{10}M_\odot$ and $j_*$ is the stellar specific angular momentum in units of $\rm 10^3 kpc~km~s^{-1}$. The maximum likelihood solution is easily computed using the {\texttt{hyper.fit}} algorithm of \citet{Robotham+2015}. This yields $p_1=0.39\pm0.04$, $p_2=-0.38\pm0.06$ and $p_3=0.06\pm0.02$ with an intrinsic scatter of standard deviation $\sigma=0.07\pm0.02$\footnote{The intrinsic Gaussian scatter of Equation~\ref{eq:master} is defined along the $\beta$-axis, but can be propagated to another set of axes appropriate to the chosen independent variable, with {consistent results}. {For example, writing Equation~\ref{eq:master} as $\log_{10}(j_*)=q_1\log_{10}(M*)+q_2\beta+q_3$ gives intrinsic scatter of $\sigma=0.20\pm0.05$ in log($j_*$), and parameters $q_1 = 1.03\pm0.11$,
$q_2 = -2.66\pm0.41$,
$q_3 = 0.17\pm0.07$, which propagate identically back to $p$.}}. The uncertainties are standard deviations, i.e.\ the square-roots of the diagonal elements of the covariance matrix, approximated as the negative inverse Hessian matrix of the likelihood at its maximum (Laplace approximation). The parameters are in agreement with OG14, while the scatter is increased due to revised $\beta$.

\begin{figure*}
\plotone{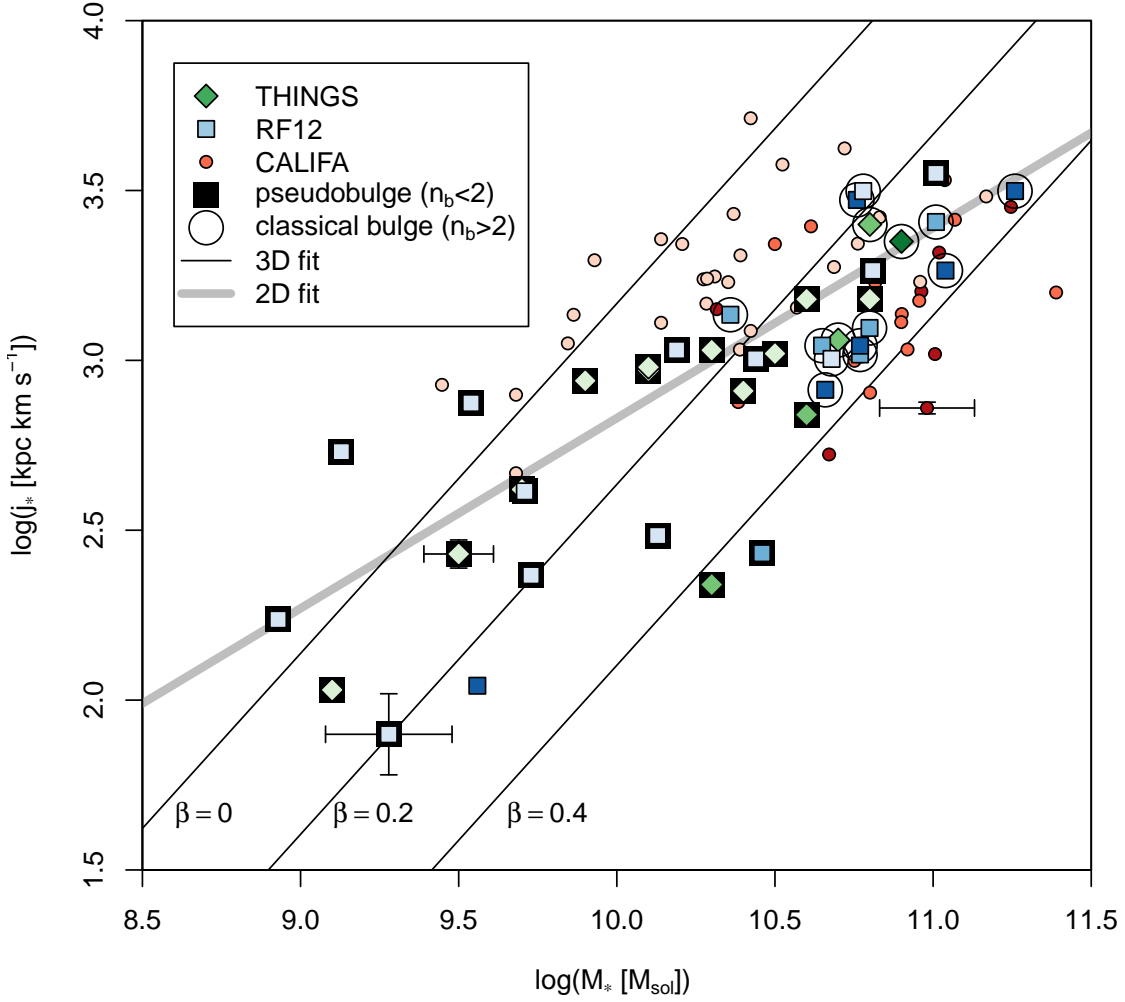}
\caption{The $M_*-j_*$ plane. 
THINGS and RF12 symbols are as in Figure~\ref{fig:betabeta}, and CALIFA as orange filled circles. Points are {assigned discrete colour shades according to bins} of $\beta ~\epsilon~ [0, 0.2, 0.4, 1)$. Black filled squares are placed around the galaxies that contain pseudobulges, and unfilled circles around those that contain classical bulges, on the basis of bulge S{\'e}rsic index. The bulge of NGC3593 is not classifiable in this manner, so is unmarked, as are the CALIFA galaxies. A typical error bar is shown for each of the THINGS, RF12 and CALIFA subsamples.
Lines of constant $\beta$ represent a trivariate fit in $M_*-j_*-\beta$ space (Equation~\ref{eqn:OG14-10}). The best-fitting exponent $\alpha = 1.03 \pm 0.11$ is consistent with $\alpha = 1$. The 2D fit (thick grey line) gives $\alpha = 0.56 \pm 0.06$, consistent with CDM predictions for halos.
\label{fig:m_j}}
\end{figure*}

In Figure~\ref{fig:m_j} we show the $M_*-j_*$ projection, overlaid with lines of constant $\beta$. These best-fitting lines take the form of eq.~(10) of OG14,  
\begin{equation}
\label{eqn:OG14-10}
\frac{j_*}{10^3\ {\rm kpc\ km\ s^{-1}}} = 
k e^{(-g\beta)} \left(\frac{M_*}{10^{10}M_\odot}\right)^\alpha,
\end{equation}
where coefficients $k$ and $g$ together are instructed by baryonic physics, $g$ modifies the bulge-dependent scale, and the exponent $\alpha$ is predicted by CDM to be $\alpha = 2/3$ for DM halos. {Equation~\ref{eqn:OG14-10} is simply obtained by potentiating Equation~\ref{eq:master}. Since our observational uncertainties tend to be normal in log($M_*$), log($j_*$) and $\beta$ (with the exception that $0\leq \beta \leq 1$), rather than in $M_*$, $j_*$ and $exp(\beta)$, it is sensible to fit the parameters ${p} = \{p_1,p_2,p_3\}$ of Equation~\ref{eq:master} and propagate them to the parameters ${q} = \{k=10^{(-p_3/p_2)},g=-ln(10)/p_2,\alpha=-p_1/p_2\}$ of Equation~\ref{eqn:OG14-10}. The covariance matrix $C_{q}$ of the new parameters ${q}$ can then be estimated by linearly propagating the covariance matrix $C_{p}$ of the parameters ${p}$, i.e.\ $C_{q} = JC_{p} J^\dag$, where the Jacobian $J$ is defined as $J_{\rm ij}=\partial{q_{\rm i}}/\partial{p_{\rm j}}$.} {This method is {consistent with} fitting Equation~\ref{eqn:OG14-10} assuming log-normal uncertainties in $M_*$ and $j_*$ and normal in $\beta$. Note that the orthogonal scatter is {not fit here but is }{also} minimized \citep[see][for details]{Robotham+2015}. } 

We give the resulting coefficients {and their uncertainties} in Table~\ref{tab:fits1}. We find $\alpha = 1.03 \pm 0.11$, consistent with OG14. {The importance of correctly accounting for measurement uncertainties in all variables as well as intrinsic scatter is demonstrated by the fact that a simple, linear least-squares fit to Equation~\ref{eqn:OG14-10} that only accounts for measurement uncertainties in $j_*$ and not intrinsic scatter yields a significantly different exponent $\alpha = 0.66 \pm 0.06$. The} bulge dependence $g$ and prefactor $k$ {from our fit to Equation~\ref{eq:master} are also} consistent {with OG14} at the 3-$\sigma$ level. The fits for THINGS or RF12 alone are consistent with the main sample; CALIFA is significantly different, but this is likely due to their angular size selection function, which results in a lack of low-mass, low-$j_*$ galaxies. Galaxies certain to contain a pseudobulge (or no bulge) are consistent with the main sample. Conversely, the galaxies in our sample that contain classical bulges {do not have sufficient dynamic range in $M_*$ to measure the presence or absence of a relation.}

The corresponding two-dimensional fit to $j_* \propto M_*^\alpha$ (that is, ignoring the effect of $\beta$) {can be found in a similar manner by fitting ${\rm log}j_* = \alpha {\rm log}(M_*) + a$; this }gives $\alpha = 0.56 \pm 0.06$, consistent with the CDM prediction for halos of $\alpha = 2/3$.

\begin{figure*}
\plotone{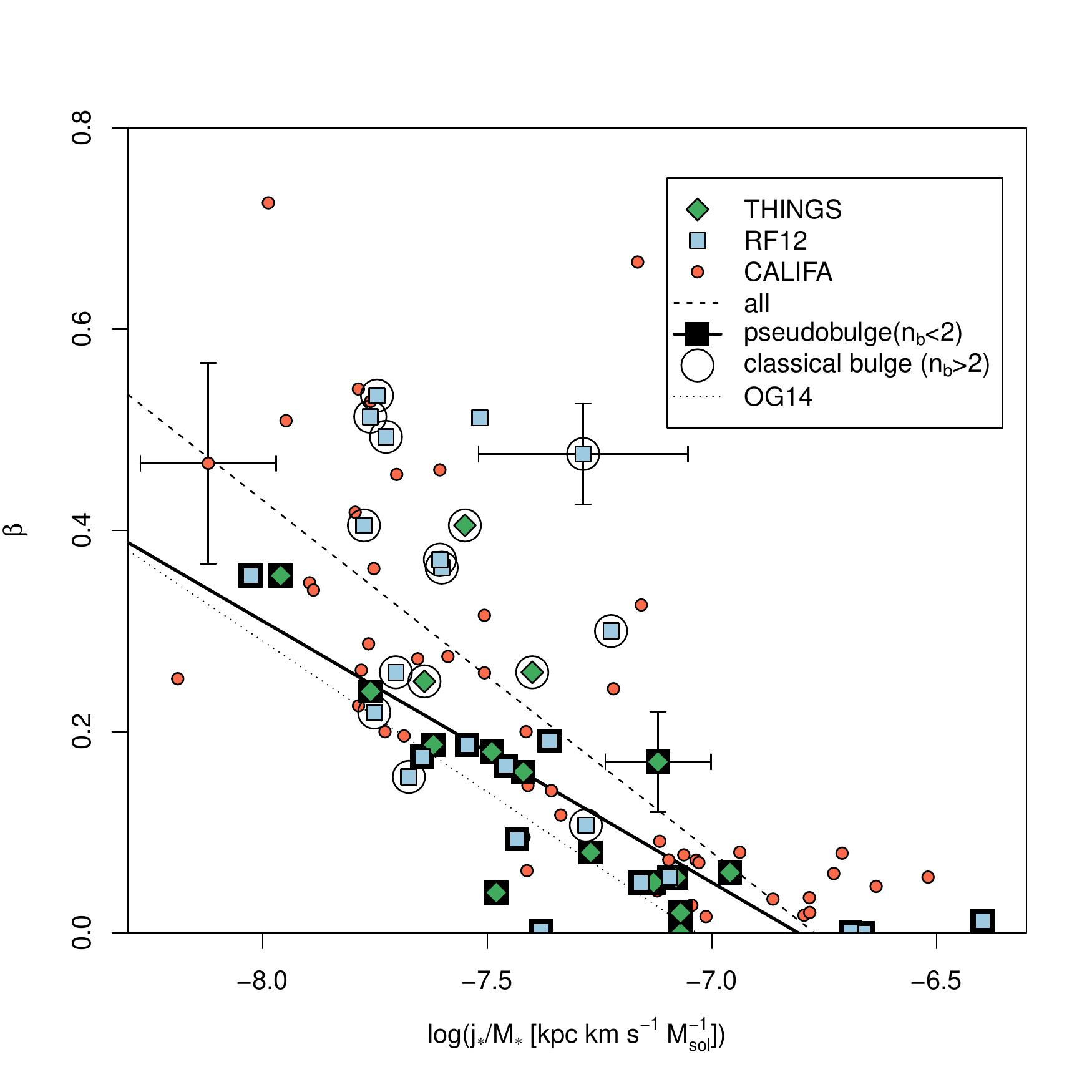}
\caption{The $\beta-j_*/M_*$ relation. Symbols are as for Figure~\ref{fig:m_j}. The solid line fits galaxies that contain a pseudobulge, the dashed line fits the entire sample, and the dotted line gives OG14. Pseudobulges appear to follow a separate sequence to classical bulges, consistent with a separate evolutionary track. 
\label{fig:bt_jm}}
\end{figure*}

\begin{deluxetable}{lLLL}
\tablecaption{Coefficients for best-fitting lines \label{tab:fits1}}
\tablecolumns{4}
\tablewidth{0pt}
\tablehead{
\colhead{$M_*-j_*$ relation (Eq.~\ref{eqn:OG14-10})} &
\colhead{$k$} &
\colhead{$g$} &
\colhead{$\alpha$}
}
\startdata
This work:\\
\-All & 1.47 \pm 0.24 & 6.13 \pm 0.95 & 1.03 \pm 0.11 \\
THINGS & 1.38 \pm 0.52 & 8.65 \pm 3.07 & 1.27 \pm 0.27 \\
RF12 & 1.75 \pm 0.82 & 6.60 \pm 2.31 & 1.07 \pm 0.21 \\
CALIFA & 1.76 \pm 0.25 & 4.25 \pm 0.79& 0.63 \pm 0.12 \\
Pseudobulge ($n_b < 2$) & 1.21 \pm 0.30 & 6.94 \pm 1.82 & 0.86 \pm 0.12 \\
\hline
OG14:\\
fixed $\alpha$ & 0.91 \pm 0.09 & 7.59 \pm 0.79 & 1.00 \\
free $\alpha$ & 0.89 \pm 0.11 & 7.03 \pm 1.35 & 0.94 \pm 0.07 \\
\hline
\multicolumn{1}{c}{$\beta-j_*/M_*$ relation (Eq.~\ref{eqn:OG14-11})} & \multicolumn{1}{c}{$k_1$} & \multicolumn{1}{c}{$k_2$} \\
\hline
This work:\\
All & -0.35 \pm 0.04 & 0.08 \pm 0.02 \\
Pseudobulge, all & -0.26 \pm 0.06 & 0.05 \pm 0.03  \\
Pseudobulge, THINGS & -0.23 \pm 0.06 & 0.04 \pm 0.02  \\
\hline
OG14 & -0.30 \pm 0.03 & -0.01 \pm 0.01  \\
\hline
\enddata
\tablecomments{Coefficients of best-fitting lines to Equations~\ref{eqn:OG14-10} and~\ref{eqn:OG14-11}. }
\end{deluxetable}

In Figure~\ref{fig:bt_jm} we fix $\alpha=1$ and show the $\beta-j_*/M_*$ relation for pseudobulges and for all bulge types. 
The best-fitting lines take the form of equation 11 of OG14: 
\begin{equation}
\label{eqn:OG14-11}
\beta = k_1{\rm log}\left(\frac{j_*\ M_*^{-1}}{10^{-7}\ {\rm kpc\ km\ s^{-1}}\ M_\odot^{-1}}\right)\ +\ k_2,
\end{equation}
with coefficients $k_1$ and $k_2$ given in Table~\ref{tab:fits1}. {This is obtained by imposing $p_2 = -p_1$ in Equation~\ref{eq:master}, and propagating parameters ${p} = \{p_1,-p_1,p_3\}$ to ${q} = \{k_1=-p_1,k_2=p_3,\alpha=-p_1/p_2=1\}$.} The fit to all bulge types has slope $k_1 = -0.35 \pm 0.04$. Galaxies with high $\beta \gtrsim 0.4$ (where most are dominated by a classical bulge) all lie above this relation, indicating a high {$\beta$ for their $j_*$ and stellar mass}.
The {sample of galaxies that host pseudobulges} follows a {shallower} relation to the fit to {galaxies with} all bulge types, with $k_1 = -0.26 \pm 0.06$, exhibiting a lower $\beta$ for a given $j_*/M_*$ than those that contain classical bulges. {Galaxies that contain classical bulges have a small range of $M_*$, so we cannot determine whether or not there is a corresponding relation for that sample. } Our sample of THINGS {galaxies with} pseudobulges is marginally consistent with {and follows a shallower relation than} OG14 (which is predominantly {comprised of galaxies that host} pseudobulges); the {main} difference is a direct result of our revised $\beta$.

\section{Discussion \& conclusion\label{sec:conclusion}}
In Section~\ref{sec:results} we presented the relation between specific angular momentum, stellar mass and bulge-to-total mass ratio. {Galaxies with} pseudobulges have lower $\beta$ per $j_*/M_*$ than {those with classical bulges}, and exhibit a well-defined track in $\beta-j_*/M_*$ space.

We investigate over what range in $\beta$ the $\beta-j_*/M_*$ relation applies. Figure~\ref{fig:bt_jm} illustrates that the relation for all galaxies extends to $\beta$ $\sim 0.4$, albeit with some scatter, partly arising from the difficulty in measuring $\beta$. As discussed in Section~\ref{sec:methods}, we employ bulge-disk decompositions for the three datasets presented here. The methods for RF12 and OG14 are identical, but the CALIFA decompositions may subtly differ due to the different imaging and software used.
There may be also systematic differences between {samples owing} to the type of kinematics used in this study: {stellar for CALIFA and RF12, and HI for THINGS}. Firstly, we have assumed (as in OG14) that the HI and stars co-rotate, but asymmetric drift may be appreciable, contributing to {$j_{gas}$ being \emph{higher} than $j_{*}$} by 0.1 dex \citep{Cortese+2016}. Those $j_{gas}$ were measured from H$\alpha$ kinematics, not HI, so we cannot apply the magnitude of that correction, but the direction of the effect will likewise be in the opposite direction to that required. Secondly, stellar kinematics trace both disk and bulge, but HI kinematics trace only the disk. {Inclusion of the bulge component in the THINGS measurements would serve to increase $j_*$ towards the CALIFA and RF12 values}, but only at an average of 0.3\% and maximum 1.3\% (OG14). Neither effect {explains why} many of the CALIFA and RF12 galaxies {with $\beta \gtrsim 0.4$ } lie significantly above the relation defined by all three samples. The apparent upturn may seem to suggest that the true relation {takes a different functional form than that assumed here.}

In general, mechanisms that increase bulge mass appear to decrease the ratio of $j_*/M_*$. 
The relation first presented in OG14 for $0 \leqslant \beta \leqslant 0.32$ is confirmed here for $0 \leqslant \beta \lesssim 0.4$, but breaks down for $\beta \gtrsim 0.4$. {At fixed $j_*/M_*$, galaxies hosting a classical bulge exhibit a range of $\beta$, extending upwards from the relation defined by galaxies that host a pseudobulge. A large $\beta$ for classical galaxies is well explained by noting from Figure~\ref{fig:m_j} that galaxies that host classical bulges are generally more massive than those that contain pseudobulges \citep[and see][]{fisher2016}, and that more massive galaxies typically have larger $\beta$ \citep{Koda+2009}. }  
The same general trend is also seen in the Evolution and Assembly of GaLaxies and their Environments simulations \citep[EAGLE,][]{Schaye+2015,Lagos+2017}, {where the most bulge-dominated galaxies lie above the best-fitting line. This is interpreted as those galaxies having higher $j_*$ than predicted by the relation, which points to an 
absence of gas-poor mergers \citep{Lagos+2017}.}

Interestingly, $ j_* \propto M_*^{\alpha}$ is well fit by $\alpha = 1$ for fixed $\beta$ over all bulge types, in line with the finding of OG14 for their sample of predominantly {galaxies with} small pseudobulges. {OG14 outlined a theoretical argument for a physical motivation to an exponent $\alpha = 1$ for a given morphology, whereby} $j_*/M_*$ traces inverse surface density, which is inversely related to the \citet{Toomre1964} Q parameter. Q quantifies instability against rotation, required for pseudobulge formation, so decreases while $\beta$ increases. {Testing this interpretation is outside of the scope of this paper but would make interesting future work.}

When ignoring the $\beta$ dimension we find $\alpha \sim 2/3$, as seen by \citet{Fall1983}, OG14 and \citet{Cortese+2016}. {This confirmation is notwithstanding several important distinctions, namely: a different bulge decomposition method to \citet{Fall1983}, OG14 and \citet{Cortese+2016}, the non-linear correction applied to the \citet{Fall1983} $j_*$ to approximate our IFU data, the use of data out to 3$r_e$ \emph{cf.} 1$r_e$ in \citet{Cortese+2016}, and the extended range of morphology with respect to OG14.} This exponent is consistent with the CDM prediction for halos $ j_{h} \propto M_{h}^{2/3}$. Connecting that prediction with our observed relation for stars implies that $M_*$ and $j_*$ depend respectively on $M_{h}$ and $j_{h}$ with the same functional form. The $M_*-M_{h}$ relation is shown to be complex \citep{Guo+2010}; future large IFS surveys \citep[e.g. Hector,][]{Lawrence+2012} are required to test whether the $j_*-j_h$ relation takes a similar form.

We investigate the possibility of two $\beta-j_*/M_*$ tracks: one for galaxies with classical bulges, thought to form by merging; and the other for galaxies with pseudobulges, formed by secular evolution \citep{kormendy2004,Wyse+1997}. 
Secular evolution refers to {angular momentum transport causing some disk material to fall towards the galaxy centre, contributing to the pseudobulge with a small increase in $\beta$. The same process  feeds star formation in the pseudobulge, which causes a small amount of $j_*$ to be lost in outflows due to stellar winds. There is a corresponding small} change in $M_*$, so the galaxy moves along a well-defined track in $\beta-j_*/M_*$. This is consistent with the distinct relation we find for pseudobulges. 
The lower $\beta$ implies that the processes that form pseudobulges are less efficient at rearranging $j_*$ and $M_*$ while forming bulges than those responsible for classical bulges. 

Conversely, mergers can significantly increase both $M_*$ and $j_*$, though some $j_*$ will cancel due to misalignment of the galaxy spin axes \citep{RF12,Lagos+2017}. There is a correspondingly large increase in $\beta$, so mergers move a galaxy \emph{above} the pseudobulge $\beta-j_*/M_*$ relation, while forming a classical bulge.
{We cannot include EAGLE as a control sample on Figure~\ref{fig:bt_jm}, since the EAGLE $\beta$ are measured from kinematic bulge-disk decompositions instead of photometric methods. As a result, they are expected to be on average larger than our $\beta$ by $\sim 0.5$ \citep[][]{Obreja+2016}. In addition to this systematic offset, there is considerable scatter, so that one cannot apply a simple correction factor. 
In future work (Lagos et al., in prep.) we will present `photometric' bulge-disk decompositions of the synthetic images of galaxies to obtain $\beta$ measurements that are directly comparable to observations. }
While EAGLE $\beta$ are not directly comparable with our $\beta$, we can use those simulations to make quantitative predictions of movement along the $j_*/M_*$ axis {, since mergers, which build classical bulges, appear to do so while moving the galaxy above the pseudobulge relation}. We see in Figure~\ref{fig:bt_jm} that {galaxies that host} classical bulges lie on or above the pseudobulge relation, so we use that relation to calculate a lower limit to $\Delta\beta$ for mergers that build classical bulges. \citet{Lagos+2017} predict that a typical dry (wet) minor merger\footnote{mass ratio $<1:3$} that increases $M_*$ by $\Delta M_* = 0.15$ dex will decrease (increase) $j_*$ by $\Delta j_* = - 0.15 (0.04)$ dex.
Combining the EAGLE predictions with our observed {relation for galaxies that contain} pseudobulges, we expect that minor mergers of $\Delta M_* = 0.15$ dex will increase $\beta$ by {more than} $\Delta\beta \geqslant 0.08 (0.03)$. Assuming that a bulgeless progenitor of a {galaxy with a} classical bulge begins at ${\rm log} (j_*/M_*) \sim -6.75$, then several mergers of this magnitude would be required to achieve $-7.7 \lesssim {\rm log} (j_*/M_*) \lesssim -7.2$ as we observe. Alternatively, a galaxy that already hosts a pseudobulge and lies in that $j_*/M_*$ range may only need to experience one such merger to form a classical bulge with those properties. However, it is not known where the progenitors of classical bulge galaxies lie in $\beta-j_*/M_*$ space. In reality there is a range of possible merger ratios, which increases the range of expected $\Delta \beta$, and may serve to explain the observed dispersion in $\beta-j_*/M_*$. The apparent failure of the relation for high $\beta$ may then simply reflect the difference between the pseudobulge and classical bulge regimes. We note that other physical processes such as outflows also modify $j_*$; these will be discussed further in Sweet et al. (in prep), when we present the internal distribution of $j_*$.

The dependence on bulge type that we see in the $\beta-j_*/M_*$ relation is reminiscent of the black hole mass -- galaxy velocity dispersion $M_{BH} - \sigma$ relation \citep{Ferrarese+2000,Gebhardt+2000}, where {galaxies with} classical bulges follow a well-defined relation in $M_{BH} - \sigma$ space but {those that contain} pseudobulges show no such relation, as suggested by \citet{Kormendy+2001} and shown by \citet{Hu2008} and \citet{Saglia+2016}. 
In this letter we see the opposite effect for $\beta-j_*/M_*$, where it is instead {galaxies with} pseudobulges that show the well-defined relation, and {those with} classical bulges that do not. This {is consistent with} earlier suggestions that classical bulges are sensitive to black hole evolution, while the evolution of pseudobulges is linked to the disk, which dominates the $j_*$ budget \citep{Kormendy+Ho2013}.

In conclusion, we have presented high-quality integrated specific angular momenta for a subset of CALIFA galaxies, and revisited the stellar mass -- specific angular momentum -- morphology relation for $0 \leqslant \beta \leqslant 0.73$, using galaxies from THINGS, RF12 and CALIFA. We confirm the OG14 $\beta-j_*/M_*$ relation for {galaxies that host} pseudobulges, albeit with increased scatter. The relation does not describe {galaxies with classical bulges}, in line with separate evolutionary channels for the formation of the two major bulge types. Future work will employ a large, homogeneous sample with high-quality measurements of $\beta$ and $j_*$ to mitigate selection biases and confirm the strength of this relation. The next critical stage is to understand the place of progenitors of galaxies that contain classical bulges, with a detailed study of specific angular momentum in main-sequence high-redshift galaxies.

\acknowledgments
We thank the anonymous referee for thoughtful comments, which helped to improve the paper.

This study uses data provided by the Calar Alto Legacy Integral Field Area (CALIFA) survey (http://califa.caha.es/). Based on observations collected at the Centro Astron{\'o}mico Hispano Alem{\'a}n (CAHA) at Calar Alto, operated jointly by the Max-Planck-Institut f{\"u}r Astronomie and the Instituto de Astrof{\'i}sica de Andaluc{\'i}a (CSIC).

Funding for the Sloan Digital Sky Survey IV has been provided by the Alfred P. Sloan Foundation, the U.S. Department of Energy Office of Science, and the Participating Institutions. SDSS-IV acknowledges
support and resources from the Center for High-Performance Computing at
the University of Utah. The SDSS web site is www.sdss.org.

SDSS-IV is managed by the Astrophysical Research Consortium for the 
Participating Institutions of the SDSS Collaboration including the 
Brazilian Participation Group, the Carnegie Institution for Science, 
Carnegie Mellon University, the Chilean Participation Group, the French Participation Group, Harvard-Smithsonian Center for Astrophysics, 
Instituto de Astrof\'isica de Canarias, The Johns Hopkins University, 
Kavli Institute for the Physics and Mathematics of the Universe (IPMU) / 
University of Tokyo, Lawrence Berkeley National Laboratory, 
Leibniz Institut f\"ur Astrophysik Potsdam (AIP),  
Max-Planck-Institut f\"ur Astronomie (MPIA Heidelberg), 
Max-Planck-Institut f\"ur Astrophysik (MPA Garching), 
Max-Planck-Institut f\"ur Extraterrestrische Physik (MPE), 
National Astronomical Observatories of China, New Mexico State University, 
New York University, University of Notre Dame, 
Observat\'ario Nacional / MCTI, The Ohio State University, 
Pennsylvania State University, Shanghai Astronomical Observatory, 
United Kingdom Participation Group,
Universidad Nacional Aut\'onoma de M\'exico, University of Arizona, 
University of Colorado Boulder, University of Oxford, University of Portsmouth, 
University of Utah, University of Virginia, University of Washington, University of Wisconsin, 
Vanderbilt University, and Yale University.

\bibliographystyle{yahapj}
\bibliography{references}

\end{document}